\documentclass[prl,aps,amssym,nofootinbib,floatfix,twocolumn,notitlepage]{revtex4-1} 

 \usepackage{xcolor}
\usepackage{amsmath,amssymb,bbold,bm}
\usepackage{graphicx}
\usepackage{caption}
\usepackage{subcaption}
\usepackage{cancel}
\usepackage{comment}

\newcommand{\be}{\begin{equation}}
\newcommand{\ben}{\begin{equation*}}
\newcommand{\ee}{\end{equation}}
\newcommand{\een}{\end{equation*}}
\newcommand{\bs}{\begin{split}}
\newcommand{\es}{\end{split}}
\newcommand{\bmx}{\begin{array}}
\newcommand{\emx}{\end{array}}
\newcommand{\bea}{\begin{eqnarray}}
\newcommand{\bean}{\begin{eqnarray*}}
\newcommand{\eea}{\end{eqnarray}}
\newcommand{\eean}{\end{eqnarray*}}
\newcommand{\dg}{^{\dagger}}
\newcommand{\dn}{^{\vphantom{\dagger}}}

\newcommand{\lr}{\leftrightarrow}

\newcommand{\ua}{\uparrow}
\newcommand{\da}{\downarrow}
\newcommand{\Ua}{\Uparrow}
\newcommand{\Da}{\Downarrow}
\newcommand{\bb}[1]{\mathbb{#1}}

\newcommand{\So}{\qquad\Rightarrow\qquad}

\newcommand{\eps}{\epsilon}

\newcommand{\tpsi}{\tilde{\psi}}

\newcommand{\sgn}[1]{{\rm sign}{#1}}
\newcommand{\pref}[1]{(\ref{#1})}

\newcommand{\re}[1]{{\rm Re}\left[ #1 \right]}
\newcommand{\im}[1]{{\rm Im}\left[ #1 \right]}
\newcommand{\tr}[1]{{\rm Tr}\Big[ #1 \Big]}

\newcommand{\abs}[1]{\left\vert #1 \right\vert}

\newcommand{\ket}[1]{\left\vert #1\right\rangle}
\newcommand{\braket}[1]{\left\langle #1\right\rangle}

\newcommand{\mat}[1]{\left(\bmx{cc}#1\emx\right)}

\newcommand{\matl}[1]{\bmx{ll}#1\emx}

\newcommand{\bw}[1]{\begin{widetext}}
\newcommand{\ew}[1]{\end{widetext}}

\newcommand{\gray}[1]{}

\newcommand{\blue}[1]{{#1}}

\begin{document}
\title{Analytical slave-spin mean-field approach to orbital selective Mott insulators }
\author{Yashar Komijani$^{1, *}$,}
\author{Gabriel Kotliar$^{1}$,}
 \affiliation{ $^1$Department of Physics and Astronomy, Rutgers University, Piscataway, New Jersey, 08854, USA}
\date{\today}
\begin{abstract}
We use the slave-spin mean-field approach to study particle-hole symmetric one- and two-band Hubbard models in presence of Hund's coupling interaction. By analytical analysis of Hamiltonian, we show that the locking of the two orbitals vs.\,orbital-selective Mott transition can be formulated within a Landau-Ginzburg framework. By applying the slave-spin mean-field to impurity problem, we are able to make a correspondence between impurity and lattice. We also consider the stability of the orbital selective Mott phase to the hybridization between the orbitals {and study the limitations of the slave-spin method for treating inter-orbital tunnellings in the case of multi-orbital Bethe lattices with particle-hole symmetry.}
\end{abstract}
\maketitle
\section{Introduction}
Iron-based superconductors  are the subject of intensive study  in the pursuit of high-temperature superconductivity \cite{Hosono, Johnston, Wang, Stewart, Dai, Si, Hirschfeld}
. 
These systems are interacting via Coulomb repulsion and Hund's rule coupling and they require the consideration of  multiple bands with crystal field and inter-orbital tunnelling \cite{TB1,TB2}. Early  DMFT studies,  pointed   out the importance of the corrlations \cite{Haule0} and Hund's rule coupling \cite{Haule1}, and reported a noticeable tendency towards orbital differentiation, with the $d_{xy}$ orbital more localized than the rest \cite{Haule2}. They also demonstrated orbital-spin separation \cite{Yin,Aron,Stadler}. Note that the orbital differentiations has been recently observed in experiments \cite{OSMTexp}. 

\blue{Another perspective on the electron correlations in these materials is that the combination of Hubbard interaction and Hund’s coupling place them in proximity to a Mott insulator \cite{Si08} and, correspondingly, the role of the orbital physics is provided by the orbital selective Mott picture \cite{Yu13,Medici14}. Ref. \cite{Yu13} demonstrated an orbital selective Mott phase in the multi-orbital Hubbard models for such materials, in the presence of the inter-orbital kinetic tunneling. In such a phase, the wavefunction renormalization for some of the orbitals vanishes. Such a phase has been observed in angle-resolved photo-emission spectroscopy (ARPES) experiments \cite{Yi13,Yi15}. Although desirable, these effects have not been understood analytically in the past, partly due to the fact that an analytical study is difficult for realistic models. However, there are simpler models, capable of capturing part of the relevant physics, which are amenable to such analytical understanding, and  this is what we study in this paper.}



The mean-field approaches to study these problems rely on various parton constructions or slave-particle techniques. The latter include slave-bosons \cite{Barnes,Coleman}, Kotliar-Ruckenstein four-boson method \cite{KR} and its rotationally invariant version \cite{Lechermann}, slave-rotor \cite{Florens}, $Z_2$ slave-spin \cite{Medici05,Hassan,Medici16,Georgescu} and its U(1) version \cite{Yu12,Yu13}, slave spin-$1$ method \cite{Vaezi} and the $Z_2$ mod-2 	 slave-spin method \cite{Ruegg, Zitko}. For a comparison of some of these methods see Appendix A. While these methods are all equivalent in the sense that they are exact representation of the partition function if the degrees of freedom are taken into account exactly, different approximation schemes required for analytical tractability, lead to different final results and therefore they have to be tested against an unbiased method like the dynamical mean-field theory (DMFT) \cite{DMFT,Koga, Song05, Bunemann, Poteryaev, Song09, Beach, Winograd, Song15} in large dimensions or density function renormalization group (DMRG) \cite{DMRG} in one dimension.



We use the $Z_2$ slave-spin \cite{Medici05,Hassan,Medici16,Georgescu} in the following to study the orbital selectivity with and without Hund's coupling. We briefly go through the method for the sake of completeness and setting the notations. By studying the free energy analytically we develop a Landau-Ginzburg theory for the orbital selectivity. A Landau-like picture has been useful in understanding the Mott transition in infinite dimensions. Using a Landau-Ginzburg approach, we show how the interaction in the slave-spin sector tend to lock the two bands together in absence of Hund's coupling and that the Hund's coupling promotes orbital selectivity. We also apply the method to an impurity problem (finite-$U$ Anderson impurity) and its use as an impurity 
show that the slave-spin mean-field result can be understood as the DMFT solution with an slave-spin impurity solver. This puts the method in perspective by showing that the mean-field result is a subset of DMFT. \blue{Additionally, we study the effect on the orbital selective Mott phase produced by inter-orbital kinetic tunnelling and point out to some of the limitations of the slave-spin for treating such inter-orbital tunnelling in particle-hole symmetric Bethe lattices. Finally, we study study the instability of the orbital selective Mott phase by including hybridization between the two orbitals.}


\subsection{$Z_2$ Slave-spin method}
We consider the Hamiltonian $H=H_0+H_{int}$, where
\be
H_0=\sum_{\braket{ij}\alpha\beta}t_{ij}^{\alpha\beta}d\dg_{i\alpha}d\dn_{j\beta}
\ee
We must demand $t_{ij}^{\alpha\beta}=[t_{ji}^{\beta\alpha}]^*$ for this Hamiltonian to be Hermitian. Unless mentioned explicitly, $\alpha$ is a super-index that contains both spin and orbital degrees of freedom. We replace the $d$-fermions with the parton construction \cite{Medici05} 
\be
d\dg_{i\alpha}=\hat{z}_{i\alpha}f\dg_{i\alpha}, \qquad \hat{z}_{i\alpha}=\tau^x_{i\alpha}. \label{eq2}
\ee
$\tau^\mu_{i\alpha}$, $\mu=x,y,z$ are SU(2) Pauli matrices acting on an slave-spin subspace per site/spin/flavour, that is introduced to capture the occupancy of the levels. Slave-spin states $\ket{\Ua_{i\alpha}}$ and $\ket{\Da_{i\alpha}}$ correspond to occupied/unoccupied states of orbital/spin $\alpha$ at site $i$, respectively. Away from half-filling, \cite{Hassan} has shown that $\tau^x_{i\alpha}$ has to be replaced with $\tau^+_{i\alpha}/2+c_\alpha\tau^-_{i\alpha}/2$ where $c$ is a gauge degree of freedom and is determined to give the correct non-interacting result. Here, for simplicity we assume $p-h$ (particle-hole) symmetry and thus maintain the form of Eq.\,\pref{eq2}. Note that this parton construction has a $Z_2$ gauge degree of freedom $\tau^{x,y}\to -\tau^{x,y}$  and $f\to -f$, thus the name $Z_2$ slave-spin. The representation \pref{eq2} increases the size of the Hilbert space. Therefore, the constraint
\be
\qquad 2f\dg_{i\alpha}f\dn_{i\alpha}=\tau^z_{i\alpha}+1,\label{eq3}
\ee
to imposed to remove the redundancy and restrict the evolution to the physical subspace. Using Eqs.\,(\ref{eq2},\ref{eq3}) it can be shown that the standard anti-commutation relations of $d$-electron are preserved.

Plugging Eq.\,\pref{eq2} in $H_0$, and imposing the constraint (on average) via a Lagrange multiplier, we have
\be
H_0=\sum_{\braket{ij}\alpha\beta}t_{ij}^{\alpha\beta}f\dg_{i\alpha}f\dn_{j\beta}\hat z\dg_{i\alpha}\hat z\dn_{j\beta}-\lambda_{i\alpha}[f\dg_{i\alpha}f\dn_{i\alpha}-(\tau^z_{i\alpha}+1)/2]\nonumber
\ee
On a mean-field level, the transverse Ising model of slave-spins can be decoupled from fermions. The decoupling is harmless in large dimensions\,\cite{Nandkishore} as the leading operator introduced by integrating over the fermions becomes irrelevant at the critical point of the transverse Ising model. Therefore, writing $H_0\approx H_{f}+H_{0S}$, we have
\bea
H_{0S}&=&\sum_{\braket{ij}\alpha\beta}{\cal J}_{ij}^{\alpha\beta} \Big[\hat z\dg_{i\alpha}\hat z\dn_{j\beta}-{Q}_{ij}^{\alpha\beta}\Big]+\sum_\alpha\lambda_{i\alpha}\tau^z_{i\alpha}/2,\nonumber\\
H_{f}&=&\sum_{\braket{ij}\alpha\beta}\tilde{t}_{ij}^{\alpha\beta}f\dg_{i\alpha}f\dn_{j\beta}-\lambda_{i\alpha}(f\dg_{i\alpha}f\dn_{i\alpha}-1/2)\label{eq8}
\eea
where $\tilde{t}_{ij}^{\alpha\beta}=t_{ij}^{\alpha\beta}Q^{\alpha\beta}_{ij}$ with $Q^{\alpha\beta}_{ij}=\langle{\hat z\dg_{i\alpha} \hat z\dn_{j\beta}}\rangle$ is the renormalized tunnelling
and ${\cal J}_{ij}^{\alpha\beta}=t_{ij}^{\alpha\beta}\langle{f\dg_{i\alpha}f\dn_{j\beta}}\rangle$ is an Ising coupling between slave-spins. The advantage of the parton construction \pref{eq2} is that the interaction $H_{int}\{\tau\}$ can be often written only in terms of the slave-spin variables, so that $H=H_f+H_S$ and $H_S=H_{0S}+H_{int}$.


\emph{Particle-hole symmetry - }
$p-h$ symmetry on the original Hamiltonian is defined as ($n$ is a site index)
\be
d_{n\alpha}\to (-1)^nd\dg_{n\alpha}, \qquad d\dg_{n\alpha}\to(-1)^n d\dn_{n\alpha}
\ee
On a bipartite lattice, the nearest neighbor tunnelling term preserves $p-h$ symmetry, even in presence of inter-orbital tunnelling. So, if the system is at half-filling the Hamitonian is invariant under $p-h$ symmetry. We have to decide what $p-h$ symmetry does to our slave-spin fields. We choose
\be
f_{n\alpha}\to(-1)^{n}f\dg_{n\alpha}, \quad \tau^x_{n\alpha}\to \tau^x_{n\alpha}, \quad \tau^z_{n\alpha}\to -\tau^z_{n\alpha}
\ee
So, we see that if the original Hamiltonian had $p-h$ symmetry, we necessarily have $\lambda_{i\alpha}=0$.

\emph{Single-site approximation} - The Hamiltonian $H_S$ is a multi-flavour transverse Ising model which is non-trivial in general. Following \cite{Medici05,Hassan,Medici16,Georgescu,Yu12,Vaezi,Ruegg,Zitko} we do a further single-site mean-field for the Ising model, exact in the limit of large dimensions:
\be
\hat z\dg_{i\alpha} \hat z\dn_{j\beta}\approx\langle{\hat z\dg_{i\alpha}}\rangle \hat z\dg_{j\beta}+\hat z\dg_{i\alpha}\braket{\hat z_{j\beta}}-\langle{\hat z\dg_{i\alpha}}\rangle\braket{\hat z_{j\beta}},\label{eq12}
\ee
The last term together with the second term of Eq.\,(8) contributes a $-2\sum_{\braket{ij}\alpha\beta}{\cal J}_{ij}^{\alpha\beta}Q_{ij}^{\alpha\beta}$. We define $z_{i\alpha}=\braket{\hat z_{i\alpha}}$ and $Z_{i\alpha}=\abs{z_{i\alpha}}^2$ as the wavefunction renormalization of orbital $\alpha$ at site $i$. The slave-spin Hamiltonian becomes (using the symmetry of ${\cal J}^{\alpha\beta}_{ij}$)
\be
H_{0S}=\sum_{i\alpha}(h^*_{i\alpha}\hat z_{i\alpha}+h.c.), \qquad h_{i\alpha}=\sum_{j\beta}{\cal J}_{ij}^{\alpha\beta}z_{j\beta}
\ee
In translationally invariant cases $h_{i\alpha}$ and $z_{i\alpha}$ become independent of the site index and ${\cal J}^{\alpha\beta}_{ij}$ depends on the distance between sites $i$ and $j$. Therefore, we can simply write $h_{\alpha}=\sum_\beta{\cal J}_{\alpha\beta}z_\beta$ where
\be
{\cal J}_{\alpha\beta}\equiv \sum_{(i-j)}{\cal J}^{\alpha\beta}_{(i-j)}=\sum_{j}t_{ij}^{\alpha\beta}\braket{f\dg_{i\alpha}f\dn_{j\beta}}.\nonumber
\ee
In absence of inter-orbital tunnelling, ${\cal J}$ is a diagonal matrix, corresponding to individual orbitals, where for each orbital ${\cal J}_{\alpha}=\int_{-D_\alpha}^{D_\alpha} d\eps\rho_\alpha(\eps) f(\eps)\eps$ is the average kinetic energy and depends only on bare parameters, unaffected by the renormalization factor $z$.  
For semicircular band (Bethe lattice), ${\cal J}=-0.2122D$, 
while for a $1D$ tight-binding model ${\cal J}_{1D}=-0.318D$ with $D=2t$.
Since the operator $\hat z_{\alpha}=\tau^x_{\alpha}$ is Hermitian, we can write the slave-spin Hamiltonian (for each site) as \cite{ftnote3}
\be
H_S=\sum_{\alpha}a_{\alpha}\tau^x_{\alpha}+H_{int}\label{eqHS}
\ee
where $ a_\alpha=2\sum_\beta{\cal J}_{\alpha\beta} z_\beta$ (at half-filling). The only non-trivial part of computation is the diagonalization of $H_S$. This is a $4^{M}$ dimensional matrix where $M$ is the number of orbitals. The free energy (per site) is
\bea
F&=&-\frac{1}{\beta}\sum_{nk}{\rm Tr}\log[-\bb G_f^{-1}(k,i\omega_n)]-2\sum_n{\cal J}_{\alpha\beta} z^*_\alpha z\dn_\beta\nonumber\\
&&\qquad\qquad-\frac{1}{\beta}\log\Big\{\tr{e^{-\beta H_S}}\Big\}.\label{eq12b}
\eea
Here $\beta=1/T$ is the inverse temperature and the second part comes from two constants introduced in Eqs.\,\pref{eq8} and \pref{eq12}. At zero temperature, the first term is just ${\cal J}_{\alpha\beta} z_\alpha^*z\dn_\beta$ and the last term is $E_S$ which depends on $z$ via $a$. Hence, 
\be
F=-\sum_{\alpha\beta} {\cal J}_{\alpha\beta} z^*_\alpha z\dn_\beta+E_S(\{a\}).
\ee
\section{One-band model}
In the one-band case the interaction is $H_{int}=U\sum_i\tilde{n}_{i\ua}\tilde{n}_{i\da}$ where $\tilde n_{i\sigma}\equiv n_{i\sigma}-1/2$. Representing the latter with $\tau^z_{i\sigma}/2$ and using translational symmetry we obtain $H_{int}\to (U/4)\tau^z_{\ua}\tau^z_{\da}$. Since we are in the paramagnetic phase ($a_\ua=a_\da$), only sum of the two spins $2\vec T=\vec\tau_\ua+\vec\tau_\da$ enter (the singlet decouples) and the Hamiltonain can be written as $H_S=2a T^x+\frac{U}{2}(T^z)^2-U/4$, creating a connection to the spin-1 representation of \cite{Vaezi}. Furthermore, we can form even and odd linear combinations of the empty and filled states and at the half-filling, only the even linear super-positions enters the the Hamiltonian. Thus, choosing atomic states of $H_S$ as
\be
\ket{\psi_{\pm 0}}=\frac{\ket{\Ua}\pm\ket{\Da}}{\sqrt{2}}, \quad
\ket{\psi_{\pm1}}=\frac{\ket{\Ua\Da}\pm\ket{O}}{\sqrt{2}}\label{eq12b}
\ee
with $E_{\pm0}=-U/4$ and $E_{\pm1}=U/4$, the Hamiltonian can be written as $H_S=2a\tau^x+(U/4)\tau^z$ where $\vec\tau$ are Pauli matrices acting between $\ket{\psi_{+0}}$ and $\ket{\psi_{+1}}$, i.e. it reduces to the $Z_2$ mod-2 slave-spin method \cite{Ruegg,Zitko}. In writing the states in Eq.\,\pref{eq12b} we have used a short-hand notation (also used in the next section) $\ket{\Ua_\ua\Da_\da}\to\ket{\Ua}$ and $\ket{\Da_\ua\Ua_\da}\to\ket{\Da}$, $\ket{\Ua_\ua\Ua_\da}\to\ket{\Ua\Da}$ and so on. 
The inset of Fig.\,(1b) shows a diagrammatic representation of the slave-spin Hamiltonian and two states decouple. The ground state of $H_S$ is that of a two-level system
\be
E_S=-\frac{U}{4}\sqrt{1+(4\alpha/U)^2}\label{eqEs}
\ee
with the level-repulsion $\alpha=2a$ and the zero-temperature (free) energy is given by [factor of $2_s$ due to spin]
\be
F=2_s\abs{\cal J}z^2+E_S(z)
\ee
The free energy is plotted in Fig.\,(1a) and it shows a second-order phase transition as $U$ is varied. Close to the the transition $\alpha\to0$, we can approximate $E_S\approx -2\alpha^2/U+8\alpha^4/U^3$. Writing the first term of the free energy as $+\alpha^2/8\abs{\cal J}$, we can read off the critical interaction $U_C=16\abs{\cal J}$.
Minimization of the free energy gives the  Gutzwiller projecion fomrula of Brinkman and Rice \cite{Rice}
\be
Z=\left\{\matl{1-u^2 & u<1 \\ 0 & u>1}\right.
\ee
with $u=U/U_C$ and is plotted in Fig.\,(1b). At finite temperature this procedure gives a first order transition terminating at a critical point\,\cite{Zitko}.

\begin{figure}[h!]
\includegraphics[width=\linewidth]{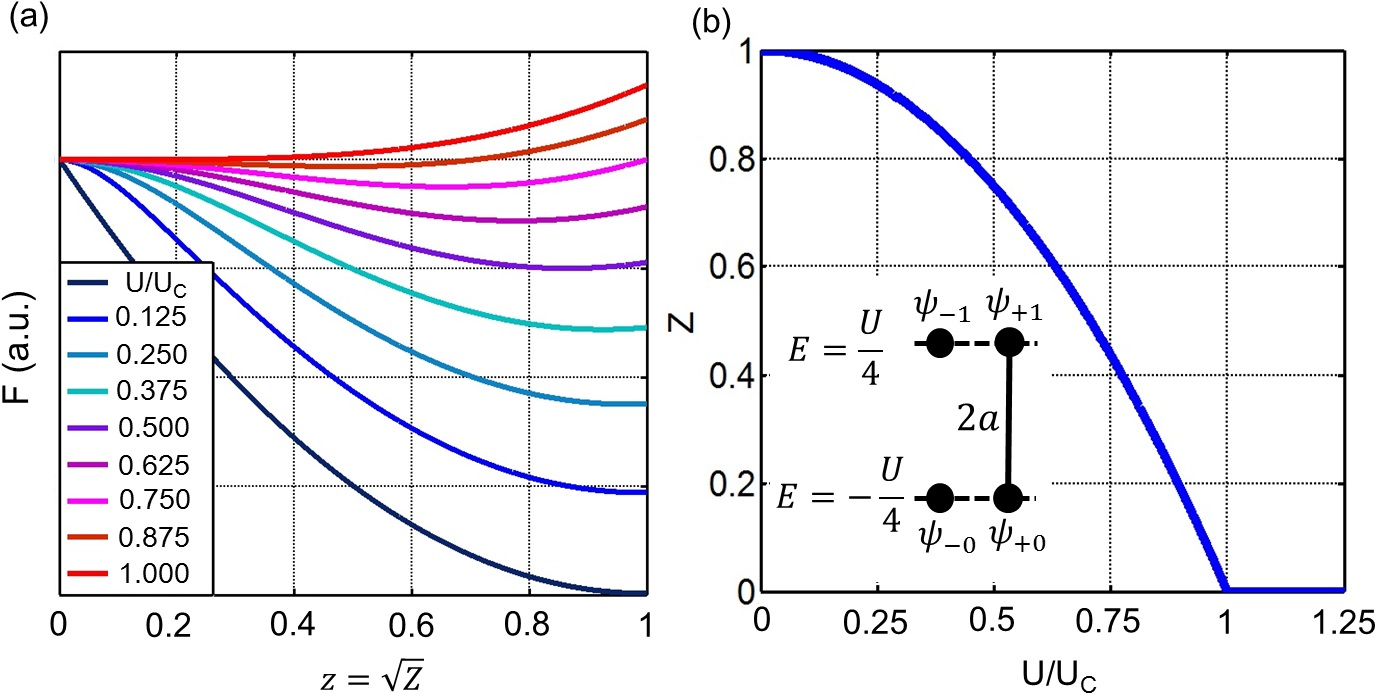}
\caption{\raggedright (color online) (a) Free energy (at $T=0$) as a function of $z$ showing a second-order phase transition as $U/U_C$ is varied. (b) Wavefunction renormalization $Z=\abs{z}^2$ as a function of $U$ has the Brinkman-Rice form. Inset: Diagrammatic representation of the slave-spin Hamiltonian. Each dot denotes on atomic state. Two states decouple and $H_S$ is equivalent to that of $Z_2$ mod-2 slave-spin.}
\end{figure}

\emph{Spectral function} - The Green's functions of the $d$-fermions $G_d(\tau)\equiv\braket{-Td_\sigma(\tau)d_\sigma\dg(0)}$ factorizes
\be
G_{d,\sigma}(\tau)\approx\braket{-Tf_{\sigma}(\tau)f_{\sigma}\dg(0)}\braket{T_\tau \tau^x_{\sigma}(\tau)\tau_{\sigma}^x(0)}
\ee
to the $f$-electron and the slave-spin susceptibility and thus the spectral function is obtained from a convolution with the slave-spin function $A_d(\omega)=A_f(\omega)*A_S(\omega)$, in which $A_f$ is a semicircular density of states with the width $Z$ and within single-site approximation $A_S$ is
\be
A_S(\omega)=Z\delta(\omega)+\frac{1-Z}{2}[\delta(\omega+2E_S)+\delta(\omega-2E_S)]
\ee
The spectral density has the correct sum-rule (in contrast to the usual slave-bosons \cite{Barnes,Coleman}) since the commutation relations of the slave-spins are preserved. However, the single-site approximation does not capture incoherent processes, and this reflects in sharp Hubbard peaks in the Mott phase ($Z=0$) where $A_f=\delta(\omega)$. Also, the spatial independence of the self-energy implies that the inverse effective-mass of ``spinons'' $m/{\tilde m}=Z[1+(m/k_F)\partial_k\Sigma]$ is zero in the Mott phase. This is again an artifact of the single-site approximation. Both of these problems are remedied, e.g. by doing a cluster mean-field calculation \cite{Hassan,Ruegg} or including quantum fluctuations around the mean-field value within a spin-wave approximation to the slave-spins\,\cite{Ruegg}. 

The fact that (beyond single-site approximation) spinons disperse in spite of $\braket{\tau^x}\to 0$ and they carry a $U(1)$ charge as seen by Eq.\,\pref{eq2}, implies that vanishing of $\braket{\tau^x}$ does not generally correspond to the Mott phase in finite dimensions. However, in large dimensions, this is correct \cite{Zitko} and that is what we refer to in the following.
\section{Two-band model}
In absence of inter-orbital tunnellings, the free-energy is 
\be
F=a_1^2/2\abs{{\cal J}_1}+a_2^2/2\abs{{\cal J}_2}+E_S(a_1,a_2)\label{eq20}
\ee
where $E_S$ is the ground state of the slave-spin Hamiltonian. For two bands we have the interaction
\bea
H_{int}&=&{U}(\tilde n_{1\ua}\tilde n_{1\da}+\tilde n_{2\ua}\tilde n_{2\da})\nonumber\\
&+&U'(\tilde n_{1\ua}\tilde n_{2\da}+\tilde n_{1\da}\tilde n_{2\ua})\nonumber\\
&+&(U'-J)(\tilde n_{1\ua}\tilde n_{2\ua}+\tilde n_{1\da}\tilde n_{2\da})+H_{XP}
\eea
where $\tilde n_\alpha \equiv n_{f\alpha}-1/2=\tau_\alpha^z/2$. The spin-flip and pair-tunnelling terms are
\bea
H_{XP}&=&-J_{X}[d\dg_{1\ua}d\dn_{1\da}d\dg_{2\da}d\dn_{2\ua}+
d\dg_{1\da}d\dn_{1\ua}d\dg_{2\ua}d\dn_{2\da}]\nonumber\\
&&+
J_P[d\dg_{1\ua}d\dg_{1\da}d\dn_{2\da}d\dn_{2\ua} +
d\dg_{2\ua}d\dg_{2\da}d\dn_{1\da}d\dn_{1\ua}
].
\eea
This term mixes the Hilbert space of $f$-electron with that of slave-spins. Following \cite{Medici05,Hassan,Medici16} we include this term approximately by $d\dg_{\alpha\sigma}\to \tau^+_{\alpha\sigma}$ and $d\dn_{\alpha\sigma}\to \tau^-_{\alpha\sigma}$ substitution so that it acts only in the slave-spin sector. The justification is that such a term captures the physics of spin-flip and pair-hopping. Using the spherical symmetry $U'=U-J$ this can be written as
\bea
H_{int}&=&\frac{U}{2}(\tilde n_{1\ua}+\tilde n_{1\da}+\tilde n_{2\ua}+\tilde n_{2\da})^2-\frac{U}{2}+H_{XP}\nonumber\\
&&-J[\tilde n_{1\ua}\tilde n_{2\da}+\tilde n_{1\da}\tilde n_{2\ua}+2\tilde n_{1\ua}\tilde n_{2\ua}+2\tilde n_{1\da}\tilde n_{2\da}]\qquad
\eea
For $J_X=J$ and $J_P=0$ it has a rotational symmetry \cite{Georges}. Alternatively, $U'=U-2J$ and $J_X=J_P=J$ has rotational symmetry. The choice does not affect the discussion qualitatively. We keep the former values in the following.\\

\emph{Atomic orbitals} - We start by diagonalizing the atomic Hamiltonian in absence of the hybridizations. Close to half-filling the doubly-occupied states have the lowest energy and are given by
\bean
&&\ket{\psi_{\pm0}}=\frac{\ket{\Ua_1\Ua_2}\pm\ket{\Da_1\Da_2}}{\sqrt{2}}, \quad E_{\pm 0}=-U-J/2,\\
&&\ket{\psi_{\pm 1}}=\frac{\ket{\Ua_1\Da_2}\pm\ket{\Da_1\Ua_2}}{\sqrt 2}, \quad E_{\pm1}=-U+J/2\mp J_X,\\
&&\ket{\psi_{\pm 2}}=\frac{\ket{\Ua\Da_1,O_2}\pm\ket{O_1\Ua\Da_2}}{\sqrt 2}, \quad E_{\pm 2}=-U+3J/2\mp J_P,
\eean
These 3 doublets become the 6-fold degenerate ground state when $J\to 0$. The $1,3$-particle states are then next
\bean
&&\ket{\psi_{\pm3}}=\ket{\Ua\Da_1}\frac{\ket{\Ua_2}\pm\ket{\Da_2}}{\sqrt 2}, \quad E_{\pm 3}=\lambda_1,\\
&&\ket{\psi_{\pm4}}=\ket{O}_1\frac{\ket{\Ua_2}\pm\ket{\Da_2}}{\sqrt 2}, \quad E_{\pm 4}=-\lambda_1\\
&&\ket{\psi_{\pm5}}=\frac{\ket{\Ua_1}\pm\ket{\Da_1}}{\sqrt 2}\ket{\Ua\Da_2}, \quad E_{\pm 5}=\lambda_2,\\
&&\ket{\psi_{\pm6}}=\frac{\ket{\Ua_1}\pm\ket{\Da_1}}{\sqrt 2}\ket{O}_2, \quad E_{\pm 5}=-\lambda_2,
\eean
and finally, there are two (empty and quadruple occupancy) states at the top of the ladder
\bean
&&\ket{\psi_{7}}=\ket{\Ua\Da}_1\ket{\Ua\Da}_2, \quad E_7=\lambda_1+\lambda_2+3U-3J/2,\\
&&\ket{\psi_{8}}=\ket{O}_1\ket{O}_2, \quad E_{8}=-\lambda_1-\lambda_2+3U-3J/2.
\eean

\begin{figure}[h!]
\includegraphics[width=1\linewidth]{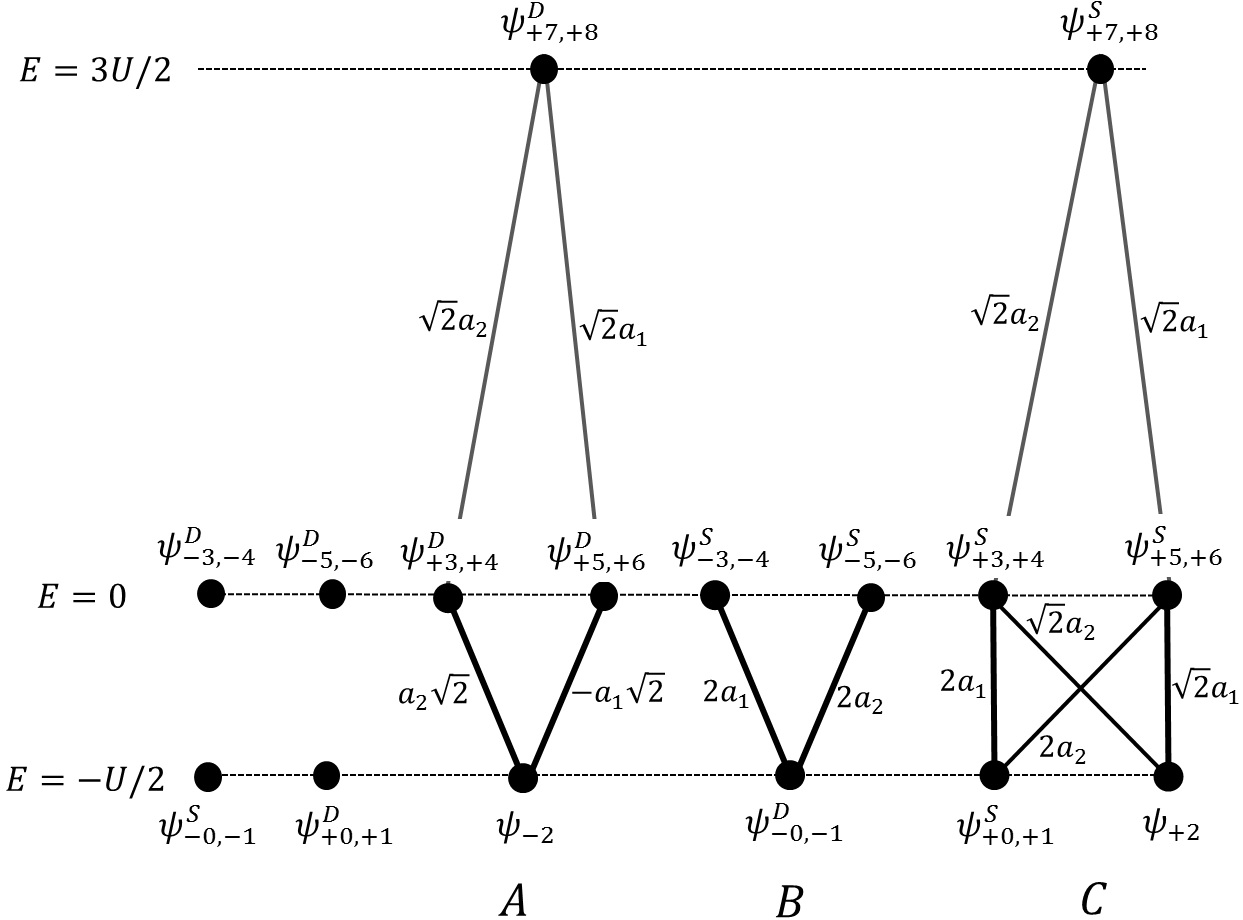}
\caption{\small\raggedright Diagrammatic representation of the slave-spin Hamiltonian $H_S$ in the two-band model with $J=0$ and $\lambda_1=\lambda_2=0$. Each dot represents an atomic state with a certain energy, denoted on the left, whereas the connecting lines represent off-diagonal elements of the Hamiltonian matrix, all assumed to be real.
We have used the short-hand notation $\sqrt{2}\psi^{S/D}_{a,b}\equiv \psi_a\pm\psi_b$. Also note that $a_i=2{\cal J}_iz_i$. The Hamiltonian factorizes into several sectors.\label{figJ0}}
\end{figure}

\emph{No Hund's rule coupling} - 
The hybridization causes transition among atomic states. In the case of no Hund's coupling we can block diagonalize $H_S$ into several sectors and diagrammatically represent it as shown in Fig.\,\pref{figJ0}. Therefore, the calculation can be reduced from 16$\times$16 to 5$\times$5. The larger the level-repulsion, the lower the ground state energy in each sector. The fact that the slave-spins decouple into several sectors brings about the possibility of possible ground-state crossings between various sectors as the parameters $a_1$ and $a_2$ are varied. Here, however, it can be shown that the sector $C$ has the lowest ground state energy for arbitrary parameters. 

\begin{figure}[h!]
\includegraphics[width=0.49\linewidth]{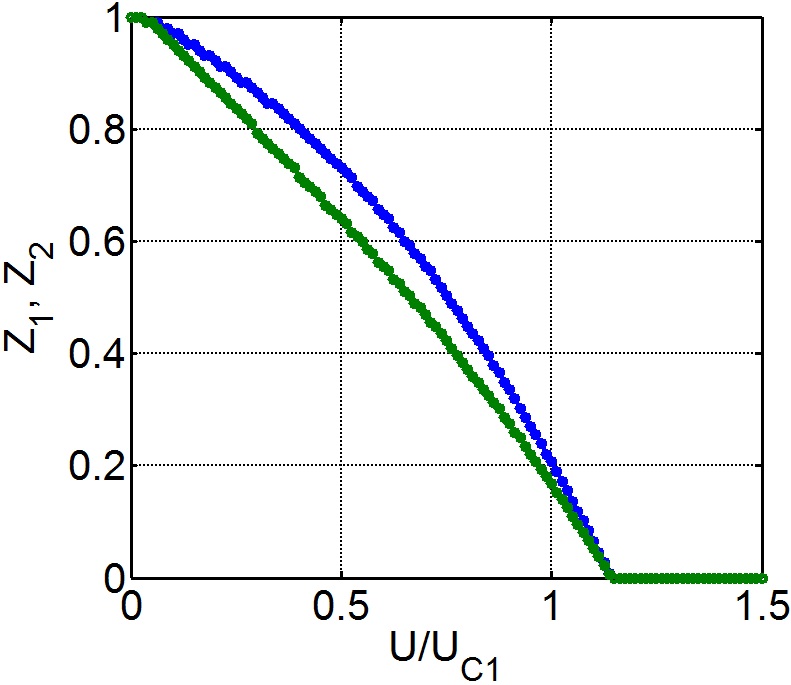}
\includegraphics[width=0.49\linewidth]{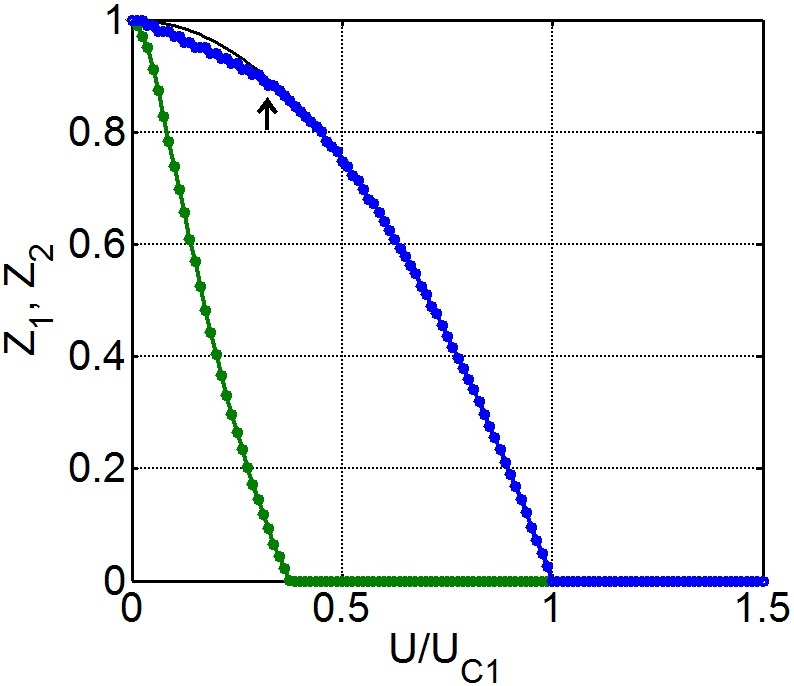}
\caption{\small\raggedright (color online Wavefunction renormalizations $Z_1$ (blue) and $Z_2$ (green) as a function of $U/U_{C1}$ in absence of Hund's rule coupling $J=0$. The states at the bottom row correspond to doubly occupied sites. The middle-row states have occupancy of 1 or 3 and the states at the top row correspond to zero or four-electron fillings. (a) Moderate bandwidth anisotropy $t_2/t_1=0.5$ shows locking. (b) Large bandwidth anisotropy $t_2/t_1=0.15$ can unlock the bands and cause OSM transition (OSMT). We also reproduce the kink in the wider-bandwidth (blue) band as the narrow band transitions to the Mott phase \cite{Medici05}, marked with an arrow. In the OSM phase, the wavefunction renormalization of the wider band follows the Brinkman-Rice formula (solid line).\label{figJ0result}}
\end{figure}

Numerical minimization of the free-energy leads to Fig.\,\pref{figJ0result} which reproduces the results of \cite{Medici05}. For $t_2/t_1>0.2$ the metal-insulator transition happens at the same critical $U$ for the two bands and we refer to it as the \emph{locking phase}, whereas for $t_2/t_1<0.2$ the critical $U$ for the bands are different $U_2<U_1$ and we refer to it as \emph{orbital selective Mott (OSM) phase}.

\begin{figure}[h!]
\includegraphics[width=\linewidth]{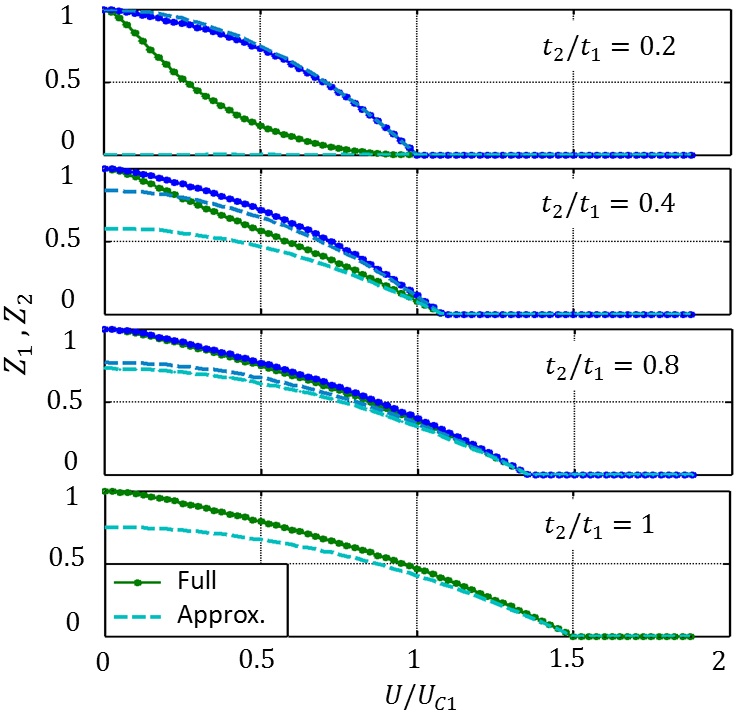}
\caption{\raggedright\small (color online) A comparison of numerical minimization of the free energy vs. the analytical two-level system. Discarding the empty and full occupancy states leads to underestimation of $Z$ as $U\to 0$ but close to the Mott transition the approximation is accurate.\label{figcomp}}
\end{figure}

In order to have the result analytically tractable we do one further simplification and that is to project out the zero and quartic occupancies per site, by dropping the high energy site at the apex of sector $C$. We expect such an approximation to be valid close to the Mott transition of the wider band, but invalid at low $U$. As a result the sector $C$ decouples into two smaller sectors $C_\pm$, each equivalent to a two-level system with the level-repulsions
\bea
\alpha_\pm &=&\sqrt{a_1^2(3/2+\sqrt 2)+a_2^2(3/2-\sqrt 2)}\nonumber\\
&&\qquad\qquad\pm\sqrt{a_1^2(3/2-\sqrt 2)+a_2^2(3/2+\sqrt 2)}.\quad\label{eqalpha}
\eea
The ground state energy of the slave-spin sector is determined with $\alpha_+$ inserted in the $E_S$ expression\,\pref{eqEs} (after an inert $-U/4$ energy shift). Note that this ground state has the $Z_2$ symmetry $a_1\lr a_2$ of the Hamiltonian $H_S$. $E_S(\alpha_+)$ as a function of $(a_1^2-a_2^2)/(a_1^2+a_2^2)$, is minimized for $a_1=a_2$. Discarding empty and filled states corresponds to truncating part of the Hilbert space and thus leads to reduced wavefunction renormalization at $U\sim 0$. In Fig.\,\pref{figcomp} we have compared our analytical solution to that of the exact result. When $a_2=0$, Eq.\,\pref{eqalpha} gives $\alpha\to 2a_1$ as in the single-band case and therefore, same critical interaction $U_{C1}=16\abs{{\cal J}_1}$ is obtained. But for symmetric bands $a_1=a_2$, it gives $\alpha=2\sqrt{3}a$. Following similar analysis as before, the free energy is $a^2/\abs{\cal J}-2\alpha^2/U$ and we obtain $U_C=24\abs{\cal J}=1.5U_{C1}$ in agreement with \cite{Medici05,Medici16}.

\emph{Locking vs.\,OSM phase} - We formulate the locking vs. OSM question as the following. Under what condition, $a_1>0$ but $a_2=0$ can be a minima of the Free energy. As mentioned before, setting $a_2=0$, $\alpha$ in Eq.\,\pref{eqalpha} reduces to the one-band $\alpha\to 2a_1$. Therefore, the Mott transition for the wide band happens at the same critical $U$ as before. To have a non-zero $a_1$ solution, we must have $U<U_{c1}$. The point $a_2=0$ always satisfies $dF/da_2=0$. To ensure that it is the energy minima we need to check the second derivative 
\be
\frac{d^2F}{da_2^2}\Big\vert_{a_2=0}=\frac{1}{\abs{{\cal J}_2}}-\frac{5}{\abs{{\cal J}_1}}>0,
\ee
which gives the condition $\abs{{\cal J}_2/{\cal J}_1}<0.2$. 

We can better understand the transition by using an order parameter. The trouble with the expression of $\alpha$ is that it cannot be Taylor expanded when $a_1$ and $a_2$ are both small. However, we may assume $a_2=ra_1$, with $r$ as an order parameter replacing $a_2$, and write down $\alpha(a_1,a_2)=a_1\alpha(r)$ where $\alpha(r)=\alpha_+(a_1\to 1,a_2\to r)$. A finite $r$ close to the transition implies locking whereas $r=0$ or $r=\infty$ implies OSM phase.  Close to the transition of both bands $\alpha\approx 0$ and we can write $E_S\approx-2\alpha^2/U+8\alpha^4/U^3$ and Eq.\,\pref{eq20} becomes
\bea
F(a_1,r)&=&\frac{a_1^2W}{2\abs{{\cal J}_1}}+O(a^4), \quad
W_x(r,u)=1+x{r^2}-\frac{\alpha^2(r)}{4u}\nonumber
\eea
Here, $x=\abs{{\cal J}_1/{\cal J}_2}$, and $u=U/U_{C1}$. The metal-insulator transition for $a_1$ happens when the mass coefficient $W$ changes sign. For negative $W$,  $a_1^2>0$ and we still have to minimize the free energy with respect to $r$. At small $r$, we can expand $\alpha(r)\approx 2+5r^2$. To zeroth order in $r$, the $W$-sign-change happen at $u=1$. 
Another transition from $r=0$ to $r>0$ happens when the corresponding mass term $(x-5/u)r^2$ changes sign, giving the same critical bandwidth ratio $x_c=5$ as we had before. So we have two equations $W(r,u)=0$ and $\partial_r W(r,u)=0$. The function $W$ is plotted in the figure and the transition from locking $r>0$ to OSM phase $r=0$ are shown.

\begin{figure}[h!]
\includegraphics[width=\linewidth]{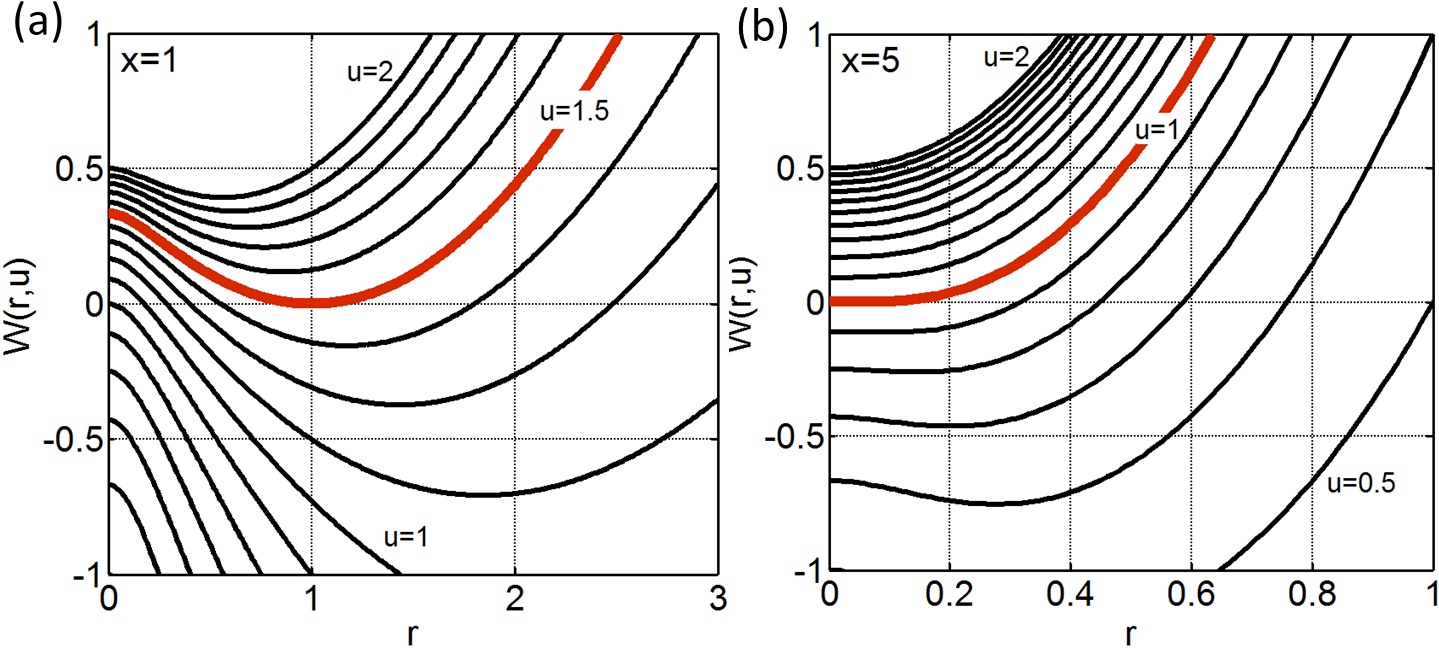}
\caption{\small \raggedright The coefficient $W(r,u)$ is shown for various $u$ as function of $r=a_2/a_1$. Equations $W=0$ and $\partial_r W=0$ are satisfied at the minimum of the red curve, which is (a) at a finite $r=1$ in the Locking phase, $\abs{{\cal J}_1}=\abs{{\cal J}_2}$. (b) and zero $r=0$ in the OSM phase, $\abs{{\cal J}_1}\ge 5\abs{{\cal J}_2}$.
}
\end{figure}

\begin{figure}[h!]
\includegraphics[width=\linewidth]{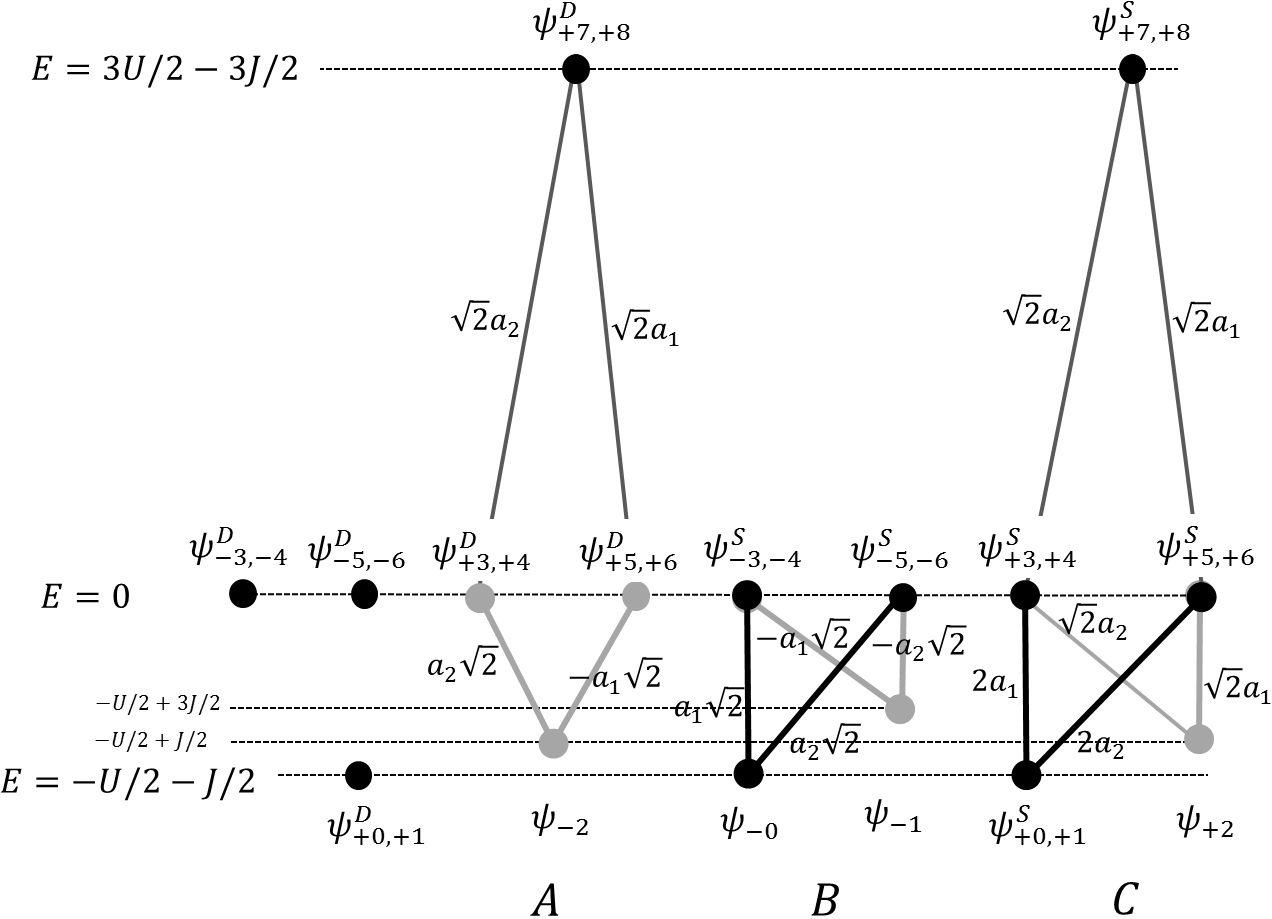}\label{fig:DiagramJ}
\caption{\small\raggedright Diagrammatic representation of the slave-spin Hamiltonian $H_S$ in the two-band model at half-filling with in presence of Hund's rule coupling $J$. Various degeneracies are lifted by $J$-interaction. In the limit of large Hund's coupling $J/U\to1/4$ we may only keep sector $C$ and neglect all the gray lines.\label{figJ}}
\end{figure}

\emph{Large Hund's coupling} - In presence of Hund's coupling the slave-spin Hamiltonian is modified to the diagram shown in Fig.\,\pref{figJ}. 

The ground state still belongs to the sector $C$. In the limit of large $J/U\to 1/4$, we may ignore all the gray lines on the block $C$ and find that the ground state is that of a two-level system, Eq.\,\pref{eqEs} with the level-repulsion
\be
\alpha=2\sqrt{a_1^2+a_2^2}
\ee
It is remarkable that the (orbital) rotational invariance of the model (even though absent in $H_S$) is recovered in this ground state. When the two bands have the same bandwidth, this formula predicts $U_C=U_{C1}$. Since $E_S$ no longer depends on $a_1^2-a_2^2$, there is no more competition between the two terms and an slight bandwidth asymmetry lead to OSM phase. This can be formulated again, following previous section, in terms of stability of a $a_1\neq 0$ but $a_2=0$ solution. We can check that 
\be
\frac{d^2F}{da_2^2}\Big\vert_{a_2=0}=\frac{1}{\abs{{\cal J}_2}}-\frac{1}{\abs{{\cal J}_1}}>0,
\ee
which gives $\abs{{\cal J}_2}<\abs{{\cal J}_1}$ as the sufficient condition for OSMT, i.e., any difference in bandwidth drives the system to the OSM phasse. Alternatively, by expanding the level-repulsion in this case $\alpha(r)\approx 2+r^2$ and plugging it into $W(r,u)$, we find that the critical bandwidth ratio $x_c=\abs{{\cal J}_1/{\cal J}_2}$ is equal to one.

\section{Tunnelling between the orbitals}
A very interesting question is about the fate of orbital selective Mott phase upon turning on an inter-orbital tunnelling. The band in Mott insulating phase has one electron per site forming localized magnetic moment. There is a large entropy associated with this phase and it is natural to expect that it would be unstable toward possible ordering. A possible mechanism that can compete with magnetic ordering, is the Kondo screening of the insulating band by the itinerant band, leading to conduction in the former and opening a hybridization gap in the latter band (effectively a new locking effect coming from Kondo screening). Within single-site approximation, however, the form of the renormalized coupling $\tilde t_{ij}^{\alpha\beta}=z^*_{i\alpha}t_{ij}^{\alpha\beta}z\dn_{j\beta}$ implies that once an orbital goes to the Mott phase, it automatically shuts down its coupling to all the other orbitals. \blue{We speculate that this effect might be responsible for the orbital selective Mott transition solution found in \cite{Yu13}. However, it is still a valid question whether or not the critical interactions $U_C$ for a Mott transition are modified by inter-orbital tunnelling, which we explore in the following.} 

Before treating inter-orbital tunnelling, we discuss how the slave-spin method can be applied to the impurity problem, and its relation to the lattice.

\subsection*{Impurity vs. Lattice and the DMFT loop} 
We can also apply the slave-spin method to an impurity problem. In particular, we can use the slave-spin (as well as any other slave-particle) method as an impurity solver for the DMFT. We show in the following that the slave-spin mean-field result corresponds to such a DMFT solution with the corresponding slave-spin impurity solver. This puts the method on firm ground and allows comparison between various methods. 

First, consider a generic $p-h$ symmetric impurity model described by the Hamiltonian $H=H_0+H_{int}$ where
\be
H_0=-\sum_{k\alpha\beta}t_k^{\alpha\beta}(d\dg_\alpha c\dn_{k\beta}+h.c.)+\sum_{k\beta}\eps_k^{\beta}c\dg_{k\beta}c\dn_{k\beta}.
\ee
Again $\alpha,\beta$ are superindices that include both orbital and spin. We have assumed that the bath is diagonal and discarded any local `crystal field' $d\dg_1 d\dn_2$ for simplicity. In the simple case of single-orbital impurity $H_{int}=U\tilde n_{d\ua}\tilde n_{d\da}$.
Via a substitution of Eq.\,\pref{eq2}, the hybridization term becomes $
H_0=-\sum_{k\alpha\beta} t_k^{\alpha\beta}(f\dg_\alpha \tau^x_\alpha c\dn_{k\beta} +h.c.)$. This problem can be written in a similar way as before $H\approx H_{f}+H_S$ where $H_S$ is exactly what we had in single-band lattice case. However, since the $f\dg_{\alpha} \tau^x_{\alpha}c\dn_{k\beta}$ interaction happens only on the impurity site, we do not need the second single-site approximation here, and obtain $a_\alpha=-2\sum_{k\beta}^{\alpha\beta}t_k^{\alpha\beta}\langle{f\dg_{k\alpha}c\dn_{k\beta}}\rangle$. In order to have a general formalism that applies to both impurity and lattice, as well as scenarios with inter-orbital tunnelling for which ${\cal J}_{\alpha\beta}$ renormalizes and is difficult to compute, we regard $a$ and $z$ as independent variables and write the free energy of Eq.\,\pref{eq12b} as \cite{ftnote3}
\be
F(\{z,a\})=F_f(\{z\})+F_S(\{a\})-\sum_\alpha a_\alpha z_\alpha.\label{eq28}
\ee
The saddle-point of $F$ with respect to $a$ and $z$ gives the correct mean-field equations. $F_f$ is the free energy of the $f$-electron given by $F_f=-T\sum_n{\rm Tr}\log[-\bb G_f^{-1}(i\omega_n)]$ where $\bb G^{-1}_f(i\omega_n)=i\omega_n\bb 1-\bb z\dg\bb\Delta(i\omega_n)\bb z$ with $\bb\Delta(i\omega_n)=\sum_k\bb t\dg\bb G_c(k,i\omega_n)\bb t$, the hybridization function. The slave-spin part is given by $F_S={\rm Tr}[e^{-\beta H_S}]$ where for a single-orbital Anderson impurity, $H_S=2a\tau^x+U\tau^z/4$, as we had in the single-band case before.

The mean-field equations w.r.t $z$ and $a$ are, respectively
\bea
&&a_{\alpha}=\frac{1}{z_\alpha}\int{\frac{d\omega}{\pi}}f(\omega)\omega \im{G_{f}^{\alpha\alpha}(\omega+i\eta)},\label{eq29} \\
&& z_{\alpha}=\frac{dF_S}{da_\alpha}.\label{eq30}
\eea
The first equation provides a relation between $a$ and $z$ that generalizes $a_\alpha=2\sum_\beta{\cal J}_{\alpha\beta}z_\beta$ (see appendix D). Having expressions for $E_S(a)$ we can eliminate $a$ in favor of $z$, or vice versa, which is equivalent to a Legendre transformation. In the appendix C, we apply these equations to the (single-orbital) finite-$U$ Anderson impurity problem and show the `transition' to the Kondo phase as the temperature is lowered.

In a lattice, the free energy has the same form as Eq.\,\pref{eq28} with the difference that $F_f=-T\sum_{k,n}{\rm Tr}\log[-\bb G_f^{-1}(k,i\omega_n)]$ where the Green's function is $\bb G_f(k,\omega+i\eta)=[(\omega+i\eta)\bb 1-\bb z\dg\bb{E}_k \bb z]^{-1}$. It can be shown that exactly same mean-field equations are obtained if $G_f$ in Eq.\,\pref{eq29} is replaced with $G_f^{\alpha\alpha}(\omega+i\eta)\to\sum_kG_f^{\alpha\alpha}(k,\omega+i\eta)$.
  Therefore, we conclude that the two problems (lattice and impurity) are equivalent provided that the hybridization function in the impurity problem is chosen such that the impurity Green's function and the local Green's function of the lattice are equal, i.e.
\be
[i\omega_n\bb 1-\bb z\dg\bb\Delta(i\omega_n)\bb z]^{-1}=\sum_k[(\omega+i\eta)\bb 1-\bb z\dg\bb{E}_k \bb z]^{-1}.
\ee
which is the DMFT consistency equation. Therefore, slave-spin mean-field is equivalent to a DMFT solution using the slave-spin method as the impurity solver. Also, note that a lattice problem in the OSM phase, corresponds to an impurity problem in which the hybridization of one of the orbitals to the bath has been turned off \cite{Medici05b}. See also Appendix B.

\subsection{Inter-orbital tunnelling}
Slave spins have been used to study Iron-based superconductors \cite{Yu13} where the inter-orbital tunnelling are important. \blue{We study this tunnelling effect in the specific case with $p-h$ symmetry and without orbital-splitting (which allows for analytic calculations). The cases that go beyond such conditions, as arising in the models for the iron-based superconductors \cite{Yu13}, remain to be explored and are left for future work. 

A troublesome feature of the slave-spins is that they break the rotational symmetry among the orbitals. Within the $p-h$ symmetric Bethe lattices that we study here, this rotational variation leads to ambiguities in presence of inter-orbital tunnellings, as we point out here.}

Let us consider a 1D chain with two orbitals $H_0=-\sum_{n\sigma} (D\dg_{n\sigma}\bb TD_{n+1,\sigma}+h.c.)$, no Hund's coupling in $H_{int}$, and a dispersion
\be
\bb E_k=-2\bb T\cos k, \qquad \bb T=\mat{t_{11} & t_{12} \\ t_{12} & t_{22}}
\ee
We have chosen $t_{12}=t_{21}$ and all the elements real (and positive)to preserve the $p-h$ symmetry. Strictly speaking, in 1D the mean-field factorization that led to Eq.\,\pref{eq8} and the consequent single-site approximation are both unjustified. The choice of dimensionality, here, is only for the ease of discussion and not essential to the conclusions. As long as the dispersion matrix can be diagonalized with a momentum-independent unitary transformation (as well as any Bethe lattice, see the appendix D), the following discussion applies. 
Diagonalizing the tunnelling matrix gives $E_k^{\pm}=-2t^\pm\cos k$ with
\be
t^{\pm}=\frac{t_{11}+t_{22}}{2}\pm\sqrt{\Big(\frac{t_{11}+t_{22}}{2}\Big)^2-\det \bb T}
\ee
Including renormalization just changes $t_{\alpha\beta}\to \tilde t_{\alpha\beta}$.
We can simply use the diagonalized form of the tunnelling matrix to calculate $F_0$ at $T=0$. Assuming $\det\bb T>0$, 
\bean
F_f&=&\sum_{\gamma=\pm}\int\frac{dk}{2\pi}E_k^{\pm}f(E_k^{\gamma})\\
&\to&-(\tilde{t}^++\tilde{t}^-)\int_{-\pi/2}^{\pi/2}\frac{dk}{\pi}\cos(k)=-2(\tilde{t}_{11}+\tilde{t}_{22})/\pi
\eean
Note that $t_{12}$ does not enter the free energy. Inserting this expression into Eq.\,\pref{eq28} and setting $d{F}/dz_i=0$, we can remove $a_i$ in favor of $z_i$. This seems to imply that there is a finite threshold (topological stability) for inter-orbital tunnelling: as long as $\det\bb T>0$, introducing $t_{12}$ does not change anything in the problem and it simply drops out and OSM phase is stable against inter-orbtital tunnelling. For large $t_{12}$ eventually $\det \bb T<0$. So, we get $t^+>0$ and $t^-<0$ and second band is inverted and $F_0$ becomes
\be
F_f\to-2(\tilde{t}^+-\tilde{t}^-)/\pi=-\frac{4}{\pi}\sqrt{\Big(\frac{\tilde{t}_{11}-\tilde t_{22}}{2}\Big)^2+\abs{\tilde{t}_{12}}^2}
\ee
Hence, $t_{12}$ has non-trivial effects on renormalization. 

On the other hand, we could have used the rotational invariance of $H_{int}$ and done a rotation in $d_1-d_2$ basis to band-diagonalize $H_0$ with the bandwidths $\bb T\to{\rm diag}\{t^+,t^-\}$, before using slave-spins to treat the interactions. It is clear then that $t_{12}$ always has non-trivial effects by modifying $t^\pm$. For example we could start in the locking phase where $t^-/t^+>0.2$, and by increasing $t_{12}$ slightly get to the OSMT phase $t^-/t^+<0.2$, without changing the sign of $\det\bb T$. This paradox exist for any $p-h$ symmetric lattice with diagonalizable tunnelling matrix. The root of the problem is that our expression in Eq.\,\pref{eqHS} is not invariant under rotations between various orbitals. \blue{Therefore, the critical value where the OSM phase persists, is basis-dependent. 
This ambiguity calls for the use of unbiased techniques to understand the role of inter-orbital tunnelling on OSMT. It might be that the model we studied analytically here is a singular limit which can be avoided by breaking $p-h$ symmetry and inclusion of crystal field in more realistic settings \cite{Yu13}. This remains to be explored in a future work.}

As discussed in \cite{Lechermann}, the way to achieve rotational-invariance is to liberate the $f$-electrons that describe quasi-particles from the physical $d$-electrons. This is achieved by a $d_\alpha\to\sum_{\beta}\hat z_{\alpha\beta}f_\beta$ representation which leads to a wavefunction-renormalization matrix $z_{\alpha\beta}=\braket{\hat z_{\alpha\beta}}$ with off-diagonal elements.  
So far, we have not been able to generalize the slave-spin to a rotationally invariant form and we leave it as a future project.

\section{On-site inter-orbital hybridization}
Even though models for the Iron-based superconductors have finite crystal level splitting and no on-site hybridization, it is interesting to introduce a hybridization between the two orbitals within the current formalism \cite{Medici05}. This is interesting, because the on-site hybridization, does not suffer from the singe-site approximation $\braket{\hat z_{i\alpha} \hat z_{i\beta}}\neq \braket{\hat z_{i\alpha}}\braket{\hat z_{i\beta}}$, as opposed to the inter-orbital tunnelling and $\braket{\hat z_{i\alpha} \hat z_{i\beta}}$ appears as an independent order parameter, which leads to the emergence of Kondo screening as we show in this section.

We can include a term $\sum_{n,\sigma}(v_{12}d\dg_{n,1\sigma}d\dn_{n,2\sigma}+h.c.)$ to the Hamiltonian. In order to preserve the $p-h$ symmetry, $v_{12}$ has to be purely imaginary. The modifications to the mean-field Hamiltonians are
\bea
\Delta H_f&=&\sum_{n,\sigma}(\tilde{v}_{12}f\dg_{n,1\sigma}f\dn_{n,2\sigma}+h.c.)-2_sA_{12}Z_{12}\\
\Delta H_S&=&\sum_{\sigma}A_{12}\tau^x_{1\sigma}\tau^x_{2\sigma}
\eea
where $\tilde v_{12}=v_{12}Z_{12}$ with $Z_{12}=\braket{\tau^x_{1\sigma}\tau^x_{2\sigma}}$ and $A_{12}=v_{12}\sum_{n}\langle{f\dg_{n,1\sigma}f\dn_{2\sigma}}\rangle+h.c.$. $Z_{12}$ and $A_{12}$ are are related to each other via the Hamiltonian above and they are independent of $\sigma$ in the paramagnetic regime. Alternatively, we can regard them as independent and impose the mean-field equation $Z_{12}=\partial F_S/\partial A_{12}$ to eliminate $A_{12}$ by a Lagrange multiplier. Assuming a small $A_{12}$ we can compute the change in slave-spin energy using second-order perturbation theory. The result is of the form $\Delta E_S=\gamma (A_{12})^2$ where $\gamma$ is (in absence of Hund's coupling) a positive constant which contains all the matrix elements and the inverse gaps $\gamma=\sum_{j\sigma\sigma'}\braket{\psi_0\vert\tau^x_{1\sigma}\tau^x_{2\sigma}\vert\psi_j}(E_j-E_0)^{-1}\braket{\psi_j\vert\tau^x_{1\sigma'}\tau^x_{2\sigma'}\vert\psi_0}$ where $E_j$ and $\ket{\psi_j}$ are the eigenvalue/states of the $H_S$ solved in the previous section. Eliminating $A_{12}$ in favor of $Z_{12}$ we find that the free energy of the system is
\bea
F(z_1,z_2,Z_{12})&=&-\frac{2_s}{\beta}\sum_{kn}{\rm Tr}\log\mat{\tilde{\eps}_{k1}-i\omega_n & iZ'_{12} \\ -iZ'_{12} & \tilde{\eps}_{k2}-i\omega_n}\nonumber\\
&&\qquad+E'_S(z_1,z_2)+\frac{(Z'_{12})^2}{\gamma'}\label{eq36}
\eea
Here $E'_S$ is the value of $E_S(a_1,a_2)-\sum_ia_iz_i$ in absence of hybridization $v_{12}$ in which $a_1$ and $a_2$ are eliminated in favor of $z_1$ and $z_2$. Also, we have redefined $\abs{v_{12}}Z_{12}\to Z'_{12}$ and $\gamma \abs{v_{12}}^2\to \gamma'$. 

Eq.\,\pref{eq36} is nothing but the free energy of a Kondo lattice at half-filling \cite{Piers} with renormalized dispersions $\tilde{\eps}_{k1}$ and $\tilde{\eps}_{k2}$. In a Kondo lattice, this form of the free energy appears using $Z'_{12}$ as the Hubbard-Stratonovitch field that decouples the Kondo coupling $\gamma'\vec S_2\cdot d\dg_1\vec\sigma d_1\dn$. Here $S_2=d_2\dg\vec\sigma d_2$ is the spin of the Mott-localized band and $\gamma'$ plays the role of the Kondo coupling. As a result of this coupling, a new energy scale $T_K\sim D\exp[-1/\gamma']$ appears, with $D\sim 2\tilde{t}_{11}$ the bandwidth of the wider band, below which the Kondo screening takes place which in the $p-h$ symmetric case gaps out both bands but away from $p-h$ symmetry mobilizes the Mott localized band. Either way, we conclude that orbital selective Mott insulating phase is unstable against hybridization between the two orbitals in agreement with \cite{Medici05}. However, even though a true selective Mottness is unstable, orbital differentiation, reflected as large difference in effective mass can exist \cite{OSMTexp}. 

\section{Conclusion}

In conclusion, we have used slave-spin mean-field method to study two-band Hubbard systems in presence of Hund's rule coupling. We have developed a Landau-Ginzburg theory of the locking vs. OSMT. We discussed the relation between slave-spins and the KR boson methods (Appendix). We have also applied the method to impurity problems and shown a correspondence between the latter and the single-site approximation of the lattice using the DMFT loop. \blue{Finally, we have discussed the limitations of the slave-spin method for multi-orbital models with both particle-hole symmetry and inter-orbital tunnelling and shown that the orbital selective Mott phase is unstable against on-site hybridization between the two orbitals.}\\

\blue{We appreciate valuable discussions with P.~Coleman, T.~Ayral, M.~Metlitski, L.~de'Medici, K.~Haule and C.-H.~Yee, and in particular, a detailed reading of the manuscript and constructive comments by Q.~Si. The authors acknowledge financial support from NSF-ONR.}\\

\blue{After completion of this manuscript, we became aware of another work \cite{Si17} which contains a Landau-Ginzburg theory of OSMT in presence of the inter-orbital tunnelling. The conclusions of the two work agrees wherever there is an overlap.}

\section{Appendix}
\subsection{A. Various slave-particle methods}
For a one band model, KR introduces four bosons and uses the representation $\hat z\dg_\sigma=P^+[p\dg_\sigma e\dn+d\dg p\dn_{-\sigma}]P^-$, where $p_\sigma\dg$, $e\dg$ and $d\dg$ are (hardcore) bosonic creation operators for $\sigma$-spinon, holon and doublon, respectively and $P^\pm$ are projectors that depend on the occupations of the bosons and are introduced to normalize the probability amplitudes over the restricted set of physical states. On the other hand, a $SU(2)$ spin-variable $\vec\tau_{\alpha}$ can be represented by two Schwinger bosons $a_{\alpha}$ and $b_{\alpha}$ satisfying the constraint $a_{\alpha}\dg a_{\alpha}+b_{\alpha}\dg b_{\alpha}=1$ (hardcore-ness), via
\be
\tau^z_{\alpha}=b\dg_{\alpha}b\dn_{\alpha}-a\dg_{\alpha}a\dn_{\alpha}, \qquad 
\tau^x_{\alpha}=a\dg_{\alpha}b\dn_{\alpha}+b\dg_{\alpha}a\dn_{\alpha}
\ee
On an operator level, the two methods have the same Hilbert space as depicted in Table \pref{table1} for the case of one orbital. Average polarization of the spin along various direction in the Bloch sphere corresponds to condensation of $a$ and $b$ bosons. 

A trouble with the slave-spin representation is that the $f$-quasi-particles carry the charge of the $d$-electron and thus the disordered phase of the slave-spins (in which the $f$-electrons still disperse beyond single-site approximation) is not a proper description of the Mott phase. As a remedy, it has been suggested \cite{Yu12} to replace $\tau^x$ in Eq.\,\pref{eq2} with $\tau^+$ and fixing the problem of non-unity $Z$ in the non-interacting case by applying fine-tuned projectors $\hat z\dg=P^+\tau^+P^-$. We note that this looks quite similar to KR.

For $M$ spinful orbitals, KR requires introducing $4^M$ bosons (only one of them occupied at a time) whereas only $2M$ slave-spins are required (each with the Hilbert space of 2). Thus the size of the two Hilbert spaces are the same $2^{2M}=4^M$.
\begin{table}
\begin{tabular}{| c|c|c |c |}
\hline
  $a\dg_\ua a\dn_\ua$ & $b\dg_\ua b\dn_\ua$ & $a\dg_\da a\dn_\da$ & $b\dg_\da b\dn_\da$ 
  \\
  \hline
  1 & 0 & 1 & 0 \\
  0 & 1 & 1 & 0 \\
  1 & 0 & 0 & 1 \\
  0 & 1 & 0 & 1\\
  \hline
\end{tabular}
\qquad
\begin{tabular}{| c|c|c |c |}
\hline
$e\dg e$ & $p_\ua\dg p_\ua\dn$  & $p_\da\dg p_\da\dn$ & $d\dg d$
  \\
  \hline
1 & 0 & 0 & 0\\
0 & 1  & 0 & 0 \\
0 & 0 & 1 & 0\\
0 & 0 & 0 & 1 \\
\hline
\end{tabular}
\caption{\small Comparison of the Schwinger boson representation of the slave-spin (left)   and Kotliar-Ruckenstein slave-bosons (right).}\label{table1}
\end{table}

\subsection{B. General low-energy considerations}
Generally for a lattice we can expand the self-energy
\be
{\bb G}_d(k,\omega)=[{\omega\bb 1-\bb{E}_k-\bb{\Sigma}_d(k,\omega)}]^{-1},
\ee
Expanding the self-energy
\be
\bb{\Sigma}_{d,lat}(k,\omega)=\bb\Sigma(0,0)+\vec{k}\cdot\partial_{\vec{k}}\bb\Sigma(0,0)+\omega\partial_\omega\bb\Sigma(0,0)+\cdots\nonumber
\ee
Within single-site approximation, the second term is zero. Denoting the third term as $\partial_\omega\bb\Sigma_d\approx 1-\bb Z^{-1}$ and assuming $\bb Z=\bb z\bb z\dg$ we can write
\be
\bb G_d(k,\omega)\approx \bb z[\omega\bb 1-\bb{z}\dg{\bb E}_k\bb  z]^{-1}\bb z\dg, 
\ee
which simply means $\bb G_f(\omega)=[\omega\bb1-\tilde{\bb E}_k]^{-1}$ and the correlation functions of the slave-particles are just decoupled $\langle{\hat z\dn_{i\alpha}(\tau) \hat z_{j\beta}\dg}\rangle\to z^*_\alpha z_\beta$ within single-site approximation also discarding any time dynamics. For the tunnelling matrix, we simply have $\tilde{\bb t}=\bb z\dg{\bb t}\bb z$. Similarly, for an impurity we have
\bea
&&i\omega_n\bb 1-\bb\Sigma_{d,imp}(i\omega_n)=\bb G_{d,imp}^{-1}(i\omega_n), \\
&&\bb G_{d,loc}(i\omega_n)=\sum_k[i\omega_n-\bb {E}_k-\bb\Sigma_{d,lat}(k,i\omega_n)]^{-1}
\eea
Denoting the interaction part of the self-energy $\bb\Sigma_{d,imp}(i\omega_n)=\bb\Delta(i\omega_n)+\bb\Sigma_{d,I}(i\omega_n)$, the DMFT approximation identifies $\bb\Sigma_{d,I}(i\omega_n)=\bb\Sigma_{d,lat}(k,i\omega_n)$. Again expanding $\bb\Sigma_{d,I}(\omega)\approx (1-\bb Z^{-1})\omega$ we have
\be
\bb G_{d,imp}(i\omega_n)=\bb z[i\omega_n\bb 1-\tilde{\bb \Delta}(i\omega_n)]^{-1}\bb z\dg, 
\ee
with $\bb{\tilde\Delta}(i\omega_n)=\bb z\dg\bb\Delta(i\omega_n)\bb z$ in agreement with $\bb G_{f,imp}(i\omega_n)=[i\omega_n\bb 1-\tilde{\bb \Delta}(i\omega_n)]^{-1}$. Using the same approximation for $\bb G_{d,loc}$ leads to
\be
\bb G_{d,loc}^{-1}(i\omega_n)\to\bb z\sum_k[i\omega_n-\bb {\tilde E}_k]^{-1}\bb z
\ee
the DMFT self-consistency loop equation is $\bb G_{f,loc}(i\omega_n)=\bb G_{f,imp}(i\omega_n)$ or 
\be
\sum_k[i\omega_n-\bb {\tilde E}_k]^{-1}=[i\omega_n\bb 1-\tilde{\bb \Delta}(i\omega_n)]^{-1}.\label{eq40}
\ee
Within the slave-spin approach there are no interactions
\be
\bb\Sigma_{f,imp}=\bb z\dg\bb\Delta(i\omega_n)\bb z, \qquad \bb\Sigma_{f,I}=0,\qquad \bb\Sigma_{f,lat}=0
\ee
and Eq.\,\pref{eq40} is satisfied as it does for any non-interacting problem.

\emph{Rotation} - Using the vector $D$ for the $d$-electrons, in presence of inter-orbital tunnelling we may sometimes be able to eliminate such inter-orbital tunnelling by a rotation to $D=\bb UD_\pm$. Since $D=\bb z F$, we assume the same rotation in the $F$-space $F=\bb UF_\pm$ (otherwise they would contain inter-orbital tunnelling) and the two $\bb z$-s are related by $\bb z=\bb U\dg\bb z_\pm\bb U$. Assuming that $\bb U$ is a SO(2) matrix, and $\bb z_\pm$ is diagonal, we find
\be
\bb z=\frac{z_++z_-}{2}\bb 1-\frac{z_+-z_-}{2}\mat{-\cos 2\alpha & \sin 2\alpha \\ \sin 2\alpha & \cos 2\alpha}
\ee
which has off-diagonal elements. Note that if one of the $z_\pm$ elements vanishes, e.g. $z_-=0$, we can factorize $\bb z$
\be
\bb z=z_+\mat{\cos\alpha \\ -\sin\alpha}\mat{\cos\alpha & -\sin\alpha}.
\ee
Then, it can be seen that $\bb Z=\bb z\bb z\dg\to z_+\bb z$ has the same form. This basically means one linear combination of $f$ electrons is decoupled (localized) and the itinerant spinon band carries characters of both $d_1$ and $d_2$ bands. This basis-dependence of the orbital Mott selectivity is again an artefact due to lack of rotational invariance.
\subsection{C. Finite-$U$ Anderson model}
\begin{figure}[h!]
\includegraphics[width=0.75\linewidth]{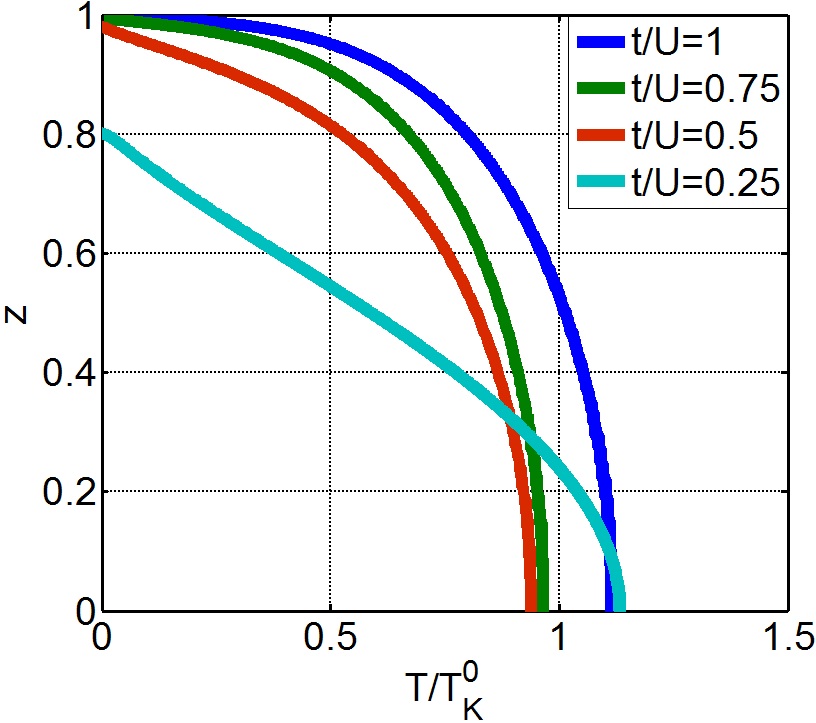}
\caption{\raggedright\small (color online) The order parameter $z$ for the Anderson model calculated from numerical evaluation of Eq.\pref{eq46}. $T_K^0=De^{-1/\rho J}$, where $J=4t^2/U$. Note that this is off with a factor of 4, an artifact of slave-spin method. We have used $D=100U$ whereas $t/U$ is varied.\label{figLast}
}
\end{figure}

The slave-spin part of the Hamiltonian is as we had in the one band case. We can use Eq.\,\pref{eq28} to eliminate $a$ in favour of $z$.
In the wide band limit for the conduction band, we have $G_f(i\omega_n)=[i\omega_n-i\Delta_K\sgn(\omega_n)]^{-1}$ where
$\Delta_K=\pi\rho t^2 z^2$, and the free energy is
\bea
F(z)&=&-{2_s}\int_{-D}^D\frac{d\omega}{\pi}f(\omega)\im{\log{(i\Delta_K-\omega)}}+E_S'(z).\qquad
\eea
$E_S'$ is obtained by eliminating $a$ from $E_S(a)-2_saz$ part of the free energy in Eqs.\,\pref{eqEs} and \pref{eq28} and is equal to $E_S'=-\frac{U}{4}\sqrt{1-z^2}$. Here, we have done a simplification to replace $F_S$ with its zero temperature value (ground state energy) while maintaining the temperature dependence of the $F_f$. We expect this approximation to be valid in the large-$U$ limit especially close to the transition. 
The mean-field equation w.r.t $z$ is
\be
z\int_{-D}^D{d\omega}f(\omega)\re{\frac {1}{\omega-i\Delta_K}}+\frac{U}{4\rho t^2}\frac{z}{\sqrt{1-z^2}}=0\label{eq43}
\ee
Close to the transition, the second term is effectively like a $z/\rho J$ with $J(z)\equiv (4t^2/U)\sqrt{1-z^2}$. 
At zero temperature the left-side simplifies
\be
z\log\frac{\Delta_K}{D}+\frac{z}{\rho J\sqrt{1-z^2}}= z\log\frac{\Delta_K}{T_K(z)}=0, 
\ee
where $T_K(z)=De^{-1/\rho J(z)}$. So to have non-zero $z$ we must have $\Delta_K=T_K$ which determines $z$. Also, we can go to non-zero temperature. We just replace the log-term in above expression with its finite-temperature expression from Eq.\,\pref{eq43}
\be
z\re{\tpsi(i\Delta_K)-\tpsi(D)}+\frac{z}{\rho J(z)}=0\label{eq46}
\ee
This is solved numerically and the result shown in Fig.\,\pref{figLast}. It shows a Kondo phase $z>0$ for $T<T_K^0$. 

\subsection{D. Stability of OSMT against interorbital tunnelling in a Bethe lattice}
Using equations of motion, the coefficient ${\cal J}_{\alpha\beta}$ defined Eq.\,\pref{eq8} can be related to the correlation function of the electrons at the same site, The result is
\be
{\cal J}^{\alpha\beta}=t^{\alpha\beta}\sum_{\eta}[\tilde t^{-1}]^{\beta\eta}\int{\frac{d\omega}{2\pi}}f(\omega)\omega A_{ii}^{\eta\alpha}(\omega)\label{eqJ}
\ee
This together with $a^{\alpha}=2\sum_\beta{\cal J}^{\alpha\beta}z^{\beta}$ leads to Eq.\,\pref{eq28}. In a Bethe lattice we can use recursive methods\,\cite{Komijani} to compute $A_{ii}^{\alpha\beta}$. When the tunnelling matrix is hermitian and there is no chemical potential or crystal field, the procedure is especially simple. We diagonalize the renormalized tunnelling matrix $\tilde{\bb t}=U \tilde {\bb t}^{D} U^{-1}$. Then the retarded and the spectral functions are\be
G^R(\omega)=\tilde{\bb t}^{-1}U\bb\Lambda(\omega) U^{-1},
\qquad
\bb A(\omega)=U \bb A^D(\omega) U^{-1}\label{eq46}
\ee
where diagonal matirix $\bb\Lambda$ contains $\lambda$-elements that satisfy $\lambda_i+\lambda_i^{-1}=\omega/t_D^i$ with the retarded boundary condition. $\bb A^D$ is diagonal matrix of semicircular density states whose width are given by the eigenvalues of $\bb{\tilde t}$. By plugging this into Eq.\,\pref{eq46} and \pref{eqJ} and using 
\be
\int{\frac{d\omega}{2\pi}}f(\omega)\omega \bb A_{ii}^D(\omega)= -0.2122\times 2\abs{\tilde{ \bb t}^D}\nonumber
\ee
we see that if the eigenvalues of the matrix $\tilde{\bb t}$ all have the same sign, then $U(\vert{\tilde{\bb t}^D}\vert=\tilde{\bb t}^D)U^{-1}=\tilde{\bb t}$. This is the generalization of the protection of OSM phase against inter-orbital tunnelling, discussed in the 1D case in the paper. For the case of two bands,
\bea
&&\det{\tilde{\bb t}}>0 \So {\cal J}^{\beta\alpha}=-0.2122\times 2 t^{\beta\alpha}\delta^{\beta\alpha},\nonumber\\
&&\det{\tilde{\bb t}}<0 \So {\cal J}^{\beta\alpha}=-0.2122\times 2 t^{\beta\alpha} R^{\alpha\beta}\qquad
\eea
i.e. for $\det\tilde{\bb t}>0$, the $\cal J$-matrix does not have any off-diagonal elements and the diagonal elements are proportional to the bare diagonal hoppings (as before), but
but if $\det\tilde{\bb t}<0$, there is a matrix $R=U\tau^zU^{-1}$ multiplying element-by-elements of  the $\cal J$-matrix which does depend on renormalization. 

Again in this problem, one could have done the rotation in $d_{\alpha}$-sector before using the slave-spins, in which case, inter-orbital tunnelling would have an effect and could cause OSM transition. \blue{Therefore, the stability found above is basis-dependent. This ambiguity is absent when $p-h$ symmetry is broken and the tunnelling matrix cannot be diagonalized independent of the momentum \cite{Yu13}.}


\begin{thebibliography}{10}
\bibitem{ftnote1}
The $Z_2$ mod-2 slave-spin introduces a spin variable that corresponds to charge per site moduli 2 (applicable to single-orbital systems), whereas the slave-spin method introduces separate slave-spin variables for each spin/orbital. To distinguish the two, we refer to the former as mod-2 slave-spin method.
\bibitem{ftnote3}
When dealing with complex $z_\alpha$, the first term of the Hamiltonian \pref{eqHS} and the last term of the free energy \pref{eq28} have to be modified to $a_\alpha z_\alpha\to(h^*_{\alpha}\hat z_{\alpha}+h.c.)$ where $h_\alpha=\sum_\beta{\cal J}_{\alpha\beta}z_\beta$ and note that $a_\alpha=h_\alpha c_\alpha+h^*_\alpha$. Since $\hat z\dg_\alpha=\hat z\dn_\alpha$ here, it is more convenient to work with $a_\alpha$.
\end{thebibliography}
\end{document}